\documentclass[prl,twocolumn,showpacs,floatfix]{revtex4}
\usepackage{amssymb}
\usepackage{amsmath}
\usepackage[dvips]{graphicx}
\usepackage{epsfig}

\def\be{\begin{eqnarray} &&}

\def\ee{\end{eqnarray}}

\def\bew{\begin{widetext}}
\def\ew{\end{widetext}}

\begin{document}

\title{Dynamical holographic QCD with area-law confinement and linear Regge
trajectories}
\author{Wayne de Paula$^{1}$, Tobias Frederico$^{1}$, Hilmar Forkel$^{1,2,3}$
and Michael Beyer$^{4}$}
\affiliation{$^{1}$Departamento de F\'{\i}sica, Instituto Tecnol\'{o}gico de Aeron\'{a}%
utica, 12228-900 S\~{a}o Jos\'{e} dos Campos, S\~{a}o Paulo, Brazil \\
$^{2}$Institut f\"{u}r Theoretische Physik, Universit\"{a}t Heidelberg,
D-69120 Heidelberg, Germany \\
$^{3}$Institut f\"{u}r Physik, Humboldt-Universit\"{a}t zu Berlin, D-12489
Berlin, Germany \\
$^{4}$Institut f\"{u}r Physik, Universit\"{a}t Rostock, D-18051 Rostock,
Germany}

\begin{abstract}
We construct a new solution of five-dimensional gravity coupled to a dilaton
which encodes essential features of holographic QCD backgrounds dynamically.
In particular, it implements linear confinement, i.e. the area law behavior
of the Wilson loop, by means of a dynamically deformed anti-de Sitter
metric. The predicted square masses of the light-flavored natural-parity
mesons and their excitations lie on linear trajectories of approximately
universal slope with respect to both radial and spin quantum numbers and are
in satisfactory agreement with experimental data.
\end{abstract}

\maketitle

%11.25.Tq    Gauge/string duality
%11.25.Wx    String and brane phenomenology
%14.40.-n    Mesons
%12.40.Yx    Hadron mass models and calculations

Over the past decade a qualitatively new perspective on strong-interaction
physics emerged from gauge/string dualities~\cite{revs} and the underlying
holographic principle. These dualities map (i)\ string theory spectra\ on
asymptotically AdS $\times X$ spacetimes (i.e. AAdS $\times X$, where $X$\
is a compact space) into gauge invariant, local operators of the dual field
theory, (ii) the fields parameterizing the boundary conditions into sources
for the dual operators, and (iii) the string theory partition function (or
its low-energy gravity limit) into the generating functional of the
field-theory correlators. As a consequence, the notoriously complex
strong-coupling regime of large-$N_{c}$ gauge theories can be approximated
(in low-curvature regions) by weakly coupled and hence analytically
treatable classical gravities. The gauge/gravity correspondence thereby
supplies new analytical tools for the study of hadronic observables in the
non-perturbative regime of the strong force.

Applications of gauge/gravity dualities to \textquotedblleft
QCD-like\textquotedblright\ gauge theories either start from specific
D-brane setups in ten- (or five-) dimensional supergravity and derive the
corresponding gauge theory properties, or try to guess a suitable background
and to improve it in bottom-up fashion by comparing the predictions to QCD
data. Even the simplest and oldest bottom-up (or \textquotedblleft
AdS/QCD\textquotedblright ) model, the hard wall \cite{hw}, reproduces a
surprising amount of hadron phenomenology \cite{hwph}. The conformal
invariance of AdS$_{5}$ in the UV reproduces, in particular, the counting
rules which govern the scaling behavior of hard QCD scattering amplitudes,
while an infrared cutoff on the fifth dimension at the QCD scale $\Lambda _{%
\text{QCD}}$ implements the mass gap and discrete hadron spectra.

The hard-wall predictions for the squared masses of light-flavor hadrons
depend quadratically on the principal and spin excitation quantum numbers 
\cite{hwph}, however, in contrast to the theoretically expected and
approximately observed linear Regge behavior \cite{ani00}. A straightforward
way to correct this shortcoming was suggested in Ref. \cite{Karch} where the
AdS$_{5}$ geometry is kept intact while an additional dilaton background
field with quadratic dependence on the extra dimension is exclusively
responsible for conformal symmetry breaking. This dilaton soft-wall model
indeed generates linear Regge trajectories $m_{n,S}^{2}\sim n+S$ for
light-flavor mesons of spin $S$ and radial excitation level $n$. (Regge
behavior can alternatively be encoded via IR deformations of the AdS$_{5}$
metric \cite{Kruczenski,FBT}.)

However, the dilaton soft wall fails a quintessential test for confining
gauge theories: the resulting vacuum expectation value (vev) of the Wilson
loop does not exhibit the area-law behavior which a linearly confining
static quark-antiquark potential would generate. This is because the Wilson
loop vev is determined by the area of the dual string world sheet in the
five-dimensional spacetime \cite{Malda2}, i.e. it depends exclusively on the
background geometry. Since the latter remains exact AdS$_{5}$ (and thus
conformal) in the soft-wall model, the Wilson loop vev shows a non-confining
perimeter law. (The hard wall, on the other hand, also confines magnetic
charges instead of screening them \cite{Kiritsis}.)\ A second, common
shortcoming of both hard- and soft-wall backgrounds is that they are not
solutions of a dual gravity. Hence their relation to the dynamics of the
sought after QCD dual remains obscure and all gauge theory vacuum properties
(including confinement, chiral symmetry breaking and condensates) have to be
imposed by hand \cite{for07}.

In the following we show how both shortcomings can be overcome, by deriving
a rather minimal AdS/QCD background which implements the area law, i.e.
linear confinement, dynamically. Our strategy will be to adopt an
IR-deformed (and hence non-conformal) AdS$_{5}$ ansatz family for the metric
which is general enough to generate the area law while keeping the fifth
dimension non-compact in order to allow for linear Regge trajectories. We
will then construct the corresponding dilaton fields and potentials such
that their combination solves the five-dimensional Einstein-dilaton
equations, and find those solutions which additionally generate linear Regge
trajectories in the highly excited meson spectrum.

To implement the above program, we start from the Einstein-Hilbert action of
five-dimensional gravity coupled to a dilaton $\Phi $, 
\begin{equation}
S=\frac{1}{2\kappa ^{2}}\int d^{5}x\sqrt{\left\vert g\right\vert }\left( -%
\emph{R}+\frac{1}{2}g^{MN}\partial _{M}\Phi \partial _{N}\Phi -V(\Phi
)\right) ,  \label{actiongd}
\end{equation}%
where $\kappa $ is the five-dimensional Newton constant and $V$ is a still
general potential for the scalar field. We then search for\ static solutions
of the corresponding field equations in which the metric is restricted to
the form 
\begin{equation}
g_{MN}=e^{-2A(z)}\eta _{MN}  \label{metric}
\end{equation}%
with $\eta =$ diag$\left( 1,-1,-1,-1,-1\right) $ and the dilaton field $\Phi
\left( z\right) $ depends on the radial coordinate only. We write the warp
factor as 
\begin{equation}
A(z)=\ln z+C\left( z\right)  \label{wf}
\end{equation}%
(our units are such that the $\mathrm{AdS}_{5}$ curvature radius is unity)
where the function $C\left( z\right) $ describes non-conformal deformations
of the AdS$_{5}$ metric. We further impose the boundary condition $C(0)=0$
which restricts the geometry to asymptotically $\mathrm{AdS}_{5}$ ($\mathrm{%
AAdS}_{5}$) spacetimes and thus ensures conformality in the ultraviolet (UV).

Variation of the action (\ref{actiongd}) leads to the Einstein-dilaton
equations 
\begin{eqnarray}
6A^{\prime }{}^{2}-\frac{1}{2}\Phi ^{\prime 2}+e^{-2A}V(\Phi ) &=&0,
\label{einsteinzz} \\
3A^{\prime \prime }-3A^{\prime }{}^{2}-\frac{1}{2}\Phi ^{\prime
2}-e^{-2A}V(\Phi ) &=&0,  \label{einstein00} \\
\Phi ^{\prime \prime }-3A^{\prime }\Phi ^{\prime }-e^{-2A}\frac{dV}{d\Phi }
&=&0  \label{dilatonequation}
\end{eqnarray}%
for the background fields $A$ and $\Phi $. This set of coupled differential
equations is redundant, i.e. only two of them are independent \cite{Kiritsis}%
. To cast those into the form most suitable for our purposes, we add the two
Einstein equations to obtain the derivative of the dilaton field as a
function of the warp factor, 
\begin{equation}
\Phi ^{\prime }=\sqrt{3}\sqrt{A^{\prime }{}^{2}+A^{\prime \prime }}~
\label{constrain}
\end{equation}%
(the positive sign is chosen for definiteness). Equation (\ref{constrain})
determines (up to a constant to be fixed by a boundary condition) the
dilaton field which forms, in combination with a given warp factor $A$, a
solution of Eqs. (\ref{einsteinzz}) -- (\ref{dilatonequation}). Equation (%
\ref{constrain}) shows, in particular, that a constant dilaton solution
requires the warp factor to satisfy $A^{\prime \prime }=-A^{\prime }{}^{2}$,
i.e. under $C\left( 0\right) =0$ exact AdS$_{5}$ is the only solution. In
other words, the conformal-symmetry breaking solutions of interest require a
non-constant dilaton and consequently an IR deformation of the AAdS$_{5}$
geometry. (The lack of the latter in the soft-wall background \cite{Karch}
explains why it cannot solve Eqs. (\ref{einsteinzz}) -- (\ref%
{dilatonequation}), as realized in Ref. \cite{Csaki}.) Another consequence
of Eq. (\ref{constrain}) is that there remains only one boundary condition
to be imposed on the dilaton field.

In order to recover the full information content of the field equations (\ref%
{einsteinzz}) -- (\ref{dilatonequation}), it remains to find the dilaton
potential $V\left( \Phi \left( z\right) \right) $ (whose specific form
played no role in the above discussion) for which a given $A$ and the
corresponding $\Phi $ determined by Eq. (\ref{constrain})\ form a solution.
This potential can be obtained either from one of the Einstein equations (%
\ref{einsteinzz}), (\ref{einstein00}) or (up to a constant) from the dilaton
equation (\ref{dilatonequation}), by substituting Eq. (\ref{constrain}). In
either case, the result is the relation \cite{Kiritsis} 
\begin{equation}
V(\Phi \left( z\right) )=\frac{3e^{2A\left( z\right) }}{2}\left[ A^{\prime
\prime }\left( z\right) -3A^{\prime }{}^{2}\left( z\right) \right]
\label{vz}
\end{equation}%
between the dilaton potential, evaluated at the dilaton solution, and the
metric. With the help of Eq. (\ref{vz}) and $\Phi ^{\prime }dV/d\Phi =dV/dz$
it can immediately be verified that \emph{all }field equations (\ref%
{einsteinzz}) -- (\ref{dilatonequation}) are indeed fulfilled by dilaton
fields which satisfy Eq. (\ref{constrain}). Hence Eqs. (\ref{constrain}) and
(\ref{vz}) are equivalent to the two independent Einstein-dilaton equations,
and they allow to obtain explicit solutions $\Phi $ and the corresponding
potential $V(\Phi \left( z\right) )$ from a given warp factor $A$ (without
requiring a superpotential).

Having shown how to construct such solutions, we can now pick a specific
warp factor suitable for our purposes. To this end, we recall that the
vacuum expectation value\ of the Wilson loop in the dual boundary theory is
related to the area of the string world sheet in the five-dimensional
space-time \cite{Malda2} and thus determined by the geometry (\ref{metric})
alone. Hence we adopt a subfamily of metrics which encodes (besides the
conformal UV behavior) the desired area law behavior. The specific
conditions given in Ref. \cite{Kinar} have been used in Ref. \cite{Kiritsis}
to show that the leading IR (i.e. $z\rightarrow \infty $) behavior of a
linearly confining (and magnetically screening) metric of the form (\ref%
{metric}) in a non-compact fifth dimension is characterized by $%
C(z)=z^{\lambda }+\dots $ where $\lambda \geq 0$ ensures that the conformal
AdS$_{5}$ metric dominates the UV limit (i.e. $C\left( 0\right) =0$) and $%
\lambda \geq 1$ is required for linear confinement.

In order to transparently elaborate on the impact of such confining $C(z)$
on the meson spectrum, we first neglect subleading terms and begin our
quantitative discussion with the minimal choice 
\begin{equation}
C_{\lambda }(z)=z^{\lambda }  \label{polinomial_metric}
\end{equation}%
for the confining AAdS$_{5}$ warp factor. We then obtain the corresponding
dilaton solution by integrating Eq. (\ref{constrain}) under the
UV-conformality preserving\ boundary condition $\Phi (0)=0$. The result is 
\begin{eqnarray}
&&\Phi _{\lambda }(z)=\frac{\sqrt{3}}{\lambda }\left[ (1+\lambda )\ln \left(
\lambda z^{\frac{\lambda }{2}}+\sqrt{\lambda +\lambda ^{2}+\lambda
^{2}z^{\lambda }}\right) \right.  \notag \\
- &&\left. (1+\lambda )\ln \left( \sqrt{\lambda +\lambda ^{2}}\right)
+z^{\lambda /2}\sqrt{\lambda +\lambda ^{2}+\lambda ^{2}z^{\lambda }}\right]
\label{phi}
\end{eqnarray}%
which, together with the AAdS$_{5}$ metric specified in Eq. (\ref%
{polinomial_metric}), forms a new analytical solution family of the
five-dimensional Einstein-dilaton equations (\ref{einsteinzz}) -- (\ref%
{dilatonequation}), with the dilaton potential given by Eq. (\ref{vz}). For $%
\lambda =2$ it reduces to a solution given in Ref. \cite{Kiritsis}. (To
elevate the soft-wall dilaton field and the AdS$_{5}$ metric to a solution
of the coupled field equations, in contrast, requires an additional tachyon
field \cite{Batell}, as suggested in Ref. \cite{Csaki}.) \ Evaluated at the
solution $\Phi _{\lambda }$, the dilaton potential is 
\begin{equation}
V\left( \Phi _{\lambda }\left( z\right) \right) =-\frac{3}{2}e^{2z^{\lambda
}}\left[ 4+7\lambda z^{\lambda }+\lambda ^{2}z^{\lambda }\left( 3z^{\lambda
}-1\right) \right] .  \label{vphiz}
\end{equation}

Close to the UV and IR limits the dilaton solution behaves as $\Phi
_{\lambda }(z)\overset{z\rightarrow 0}{\longrightarrow }c_{0}z^{\lambda /2}$
and $\Phi _{\lambda }(z)\overset{z\rightarrow \infty }{\longrightarrow }%
c_{\infty }z^{\lambda }$ where $c_{0}=2\sqrt{3(1+\lambda ^{-1})}$ and $%
c_{\infty }=\sqrt{3}$. In these limits the potential can be easily expressed
as a function of $\Phi $, i.e. $V(\Phi )\overset{\Phi \rightarrow 0}{%
\longrightarrow }-6+3(\lambda +1)(\lambda -8)\Phi ^{2}/\left(
2c_{0}^{2}\right) $ and $V(\Phi )\overset{\Phi \rightarrow \infty }{%
\longrightarrow }-9\lambda ^{2}\Phi ^{2}\exp \left( 2\Phi /c_{\infty
}\right) /\left( 2c_{\infty }^{2}\right) $. The exponential divergence for $%
\Phi \rightarrow \infty $ is both induced and counterbalanced by the
exponential behavior of the metric. Indeed, the dilaton-gravity coupling is
necessary to stabilize the classical solution since the dilaton potential is
not bounded from below. As expected, the boundary condition $\Phi (0)=0$
ensures conformal invariance in the UV where the potential $V(0)=-6$
consists of just the AdS$_{5}$ cosmological constant term.

We will now derive the mesonic excitation spectrum in the background (\ref%
{polinomial_metric}), (\ref{phi}) and then study the existence criteria for
another classic confinement signature, namely linear square mass
trajectories of Regge type for highly excited hadrons. To this end, we
utilize the tensor gauge-field framework and notation of Ref. \cite{Karch}.

The spin $S$ string modes of the massive tensor fields $\phi _{M_{1}\dots
M_{S}}$ (in axial gauge) in the dilaton-gravity background can be rewritten
in terms of reduced amplitudes $\psi _{n,S}$ which satisfy the
Sturm-Liouville equation 
\begin{equation}
\left[ -\partial _{z}^{2}+\mathcal{V}_{S}(z)\right] \psi
_{n,S}=m_{n,S}^{2}\psi _{n,S}  \label{sleq}
\end{equation}%
where the spin-dependent string-mode potential is 
\begin{equation}
\mathcal{V}_{S}(z)=\frac{B^{\prime }{}^{2}(z)}{4}-\frac{B^{\prime \prime }(z)%
}{2},  \label{vs}
\end{equation}%
with $B=\left( 2S-1\right) A+\Phi $. The gauge/gravity dictionary identifies
the eigenvalues $m_{n,S}^{2}$ with the squared meson mass spectrum of the
boundary gauge theory. Equation (\ref{vs}) implies that the mode potential $%
\mathcal{V}_{S}$ depends on $A$ and $\Phi ^{\prime }$ only, i.e. not on $%
\Phi $ and its boundary condition. Since both the dilaton solution and its
potential are determined by the confining ansatz for the warp factor $A$
(cf. Eqs. (\ref{constrain}) and (\ref{vz})) and by UV conformality, the mode
potential may also be expressed exclusively in terms of the warp factor.
Since our main focus is on the higher-spin mesons and their mass
trajectories, furthermore, we have derived the dual meson dynamics (\ref%
{sleq}) from the quadratic part of the action only, as in Ref. \cite{Karch}.
We have also neglected additional nonlinearities due to higher-dimensional
operators which would arise from stringy $\alpha ^{\prime }$ corrections in
the gravity action and in the non-Abelian DBI action of flavor brane stacks,
or from interactions with additional bulk fields. Their neglect is common
practice in the current, first generation of AdS/QCD duals. A more detailed
discussion of the ensuing limitations can be found e.g. in Refs. \cite%
{Karch,for07}.

Important qualitative aspects of the meson spectrum arising from Eq. (\ref%
{sleq}) can be understood by studying the UV (i.e. $z\rightarrow 0$) and IR (%
$z\rightarrow \infty $) limits of the mode potential. We start with the UV
behavior. Keeping only the leading terms for small $z$\ and $\lambda >1$, 
\begin{equation}
\mathcal{V}_{S}(z)=a_{0}\left( S\right) z^{-2}+a_{1}\left( S\right) z^{\frac{%
\lambda }{2}-2}+a_{2}\left( S\right) z^{\lambda -2}+\dots  \label{vsmallz}
\end{equation}%
with the spin-dependent coefficients $a_{0}\left( S\right) =S^{2}-1/4$
(which originates from the AdS$_{5}$ part of the warp factor), $a_{1}\left(
S\right) =\sqrt{3\lambda (\lambda +1)}\left( S-\lambda /4\right) $ and $%
a_{2}\left( S\right) =\lambda \left[ 8S^{2}-4(\lambda +1)S+5\lambda +3\right]
/4$. For $\lambda =4$ the dilaton field therefore becomes proportional to $%
z^{2}$ when $z\rightarrow 0$ (see the discussion below Eq. (\ref{vphiz})),
as imposed by hand in Ref. \cite{Karch}. However, while in this case the $%
a_{2}$ term in Eq. (\ref{vsmallz}) becomes harmonic and generates
approximately linear trajectories for the \emph{low}-lying radial
excitations, the mode potential will grow as $z^{6}$ at large $z$ and cause
strong nonlinearities in the trajectories of high-lying radial and spin
excitations.

To implement the linear trajectories for highly excited meson states, which
semiclassical arguments predict as a consequence of long, unbroken flux
tubes in the large-$N_{c}$ limit, we now turn to the large-$z$ behavior of
the mode potential. The leading infrared contribution is 
\begin{equation}
\mathcal{V}_{S}\left( z\right) \overset{z\rightarrow \infty }{%
\longrightarrow }\frac{\lambda ^{2}}{4}(2S+\sqrt{3}-1)^{2}\;z^{2\lambda -2}.
\label{eqn:effP}
\end{equation}%
Hence for $\lambda >1$ the entire normalizable spectrum is discrete and has
a mass gap (in the absence of normalizable zero modes), as previously found
for radial glueball excitations \cite{Kiritsis} and in remarkable agreement
with the area-law condition $\lambda \geq 1$ \cite{Kiritsis} for linear
confinement. (In contrast, the soft-wall model generates a non-confining
perimeter law \cite{Malda2}.)

\begin{figure}[tbh]
%\hspace{-2cm}
\centerline{ \epsfig{figure=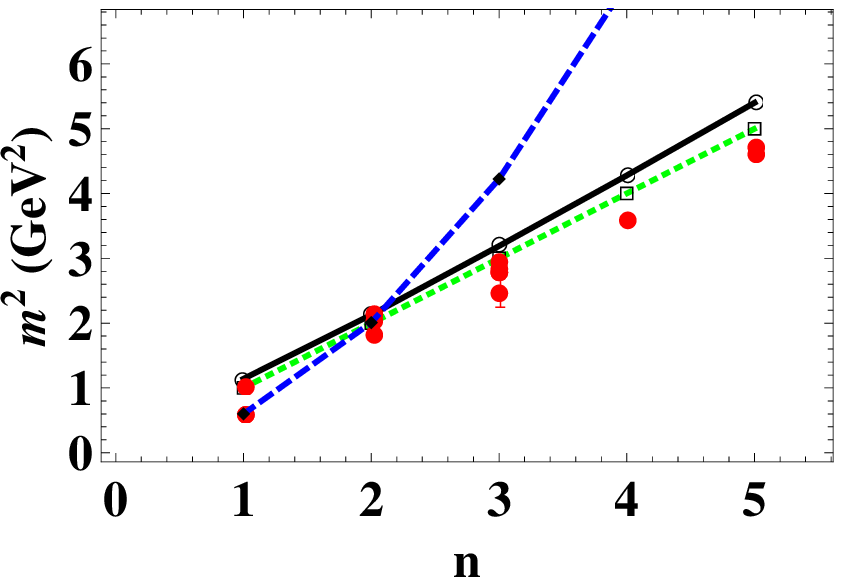,width=4cm,height=4cm}
\hspace{0.1cm} \epsfig{figure=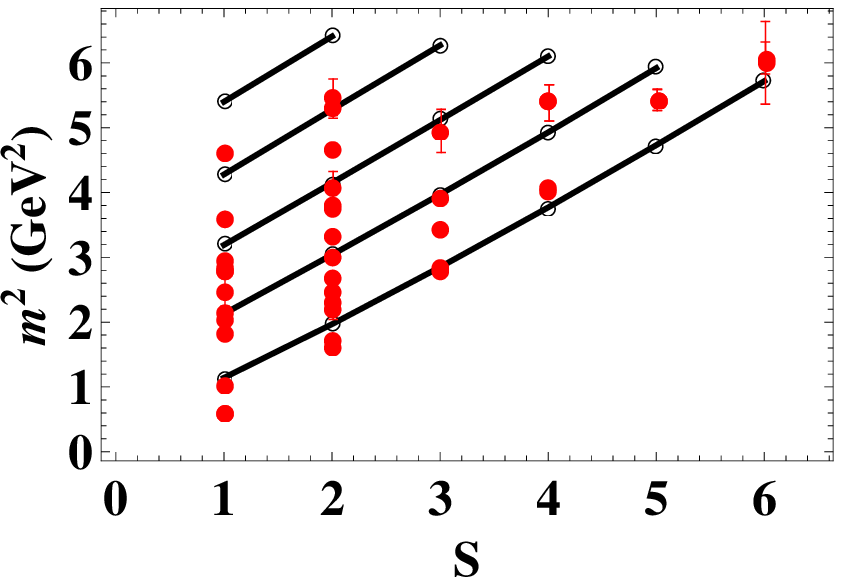,width=4cm,height=4cm}} \vspace{%
-0.5cm}
\caption{(a) Radial excitations of the rho meson in the hard-wall (dashed
line), soft-wall \protect\cite{Karch} (dotted line) and our dynamical
soft-wall (solid line, for $\Lambda _{\text{QCD}}=0.3$ GeV) backgrounds. (b)
Square mass predictions of spin excitations compared to PDG \protect\cite%
{ani00} values for $\protect\rho (770)$, $\protect\omega (782)$, $\Phi
(1020) $, $\protect\pi _{1}(1400)$, $\protect\omega (1420)$, $\protect\rho %
(1450)$, $\protect\rho (1570)$, $\protect\pi _{1}(1600)$, $\protect\omega %
(1650)$, $\Phi (1680)$ $\protect\rho (1700)$, $\protect\rho (1900)$, $%
\protect\rho (2150)$, $f_{2}(1270)$, $a_{2}(1320)$, $f_{2}(1430)$, $%
f_{2}^{\prime }(1525)$, $f_{2}(1565)$, $f_{2}(1640)$, $a_{2}(1700)$, $%
f_{2}(1810)$, $f_{2}(1910)$, $f_{2}(1950)$, $f_{2}(2010)$, $f_{2}(2150)$, $%
f_{2}(2300)$, $f_{2}(2340)$, $\protect\omega _{3}(1670)$, $\protect\rho %
_{3}(1690)$, $\Phi _{3}(1850)$, $\protect\rho _{3}(1990)$, $\protect\rho %
_{3}(2250)$, $a_{4}(2040)$, $f_{4}(2050)$, $f_{4}(2300)$, $\protect\rho %
_{5}(2350)$, $a_{6}(2450)$ and $f_{6}(2510)$.}
\label{Fig1}
\end{figure}

Asymptotically equidistant $m_{n,S}^{2}$ values corresponding to linear
trajectories further require $\lambda =2$ in the IR dominant part (\ref%
{polinomial_metric}) of the warp factor, in order to generate a harmonic IR
potential. The trajectories resulting from the simplest choice $C(z)=z^{2}$
would be of non-universal slope ($m_{n,S}^{2}\propto n\left( S+c\right) $),
however, which can be corrected by adopting 
\begin{equation}
C(z)=\frac{1+\sqrt{3}}{2S+\sqrt{3}-1}\frac{(z\Lambda _{\text{QCD}})^{2}}{%
1+e^{(1-z\Lambda _{\text{QCD}})}}  \label{cnew}
\end{equation}%
as the non-conformal warp factor. The corresponding metric remains close to
AdS$_{5}$ in the UV but deforms rather rapidly for $z\gtrsim \Lambda _{\text{%
QCD}}^{-1}$ to approach the confining large-$z$ asymptotics of Eq. (\ref%
{polinomial_metric}) with $\lambda =2$. The associated dilaton field and
potential, which turn Eqs. (\ref{wf}), (\ref{cnew}) into a solution of the
Einstein-dilaton equations (\ref{einsteinzz}) -- (\ref{dilatonequation}),
are then obtained by solving Eq. (\ref{constrain}) numerically according to
the procedure outlined above. This determines the potential (\ref{vs}), and
the masses follow by solving the eigenvalue problem (\ref{sleq})
numerically. In Fig. (\ref{Fig1}) the resulting spectrum is compared to
experimental data and hard- and soft-wall model results. Note that the state
label $n\geq 1$ is chosen such that $n=1$ refers to the nodeless ground
state of the radial excitation spectrum. A satisfactory description of the
meson mass spectrum with nearly universal Regge slopes is indeed achieved
without any tuning of adjustable parameters. (The spin dependent factor in
Eq. (\ref{cnew}) is required by universality. For a physical interpretation
see Ref. \cite{FBT}.) A good analytical approximation to the spectrum for $%
\Lambda _{\text{QCD}}=0.3$ GeV is (in units of GeV) 
\begin{equation}
m_{n,S}^{2}\simeq \frac{1}{10}\left( 11n+9S-9\right) ,\text{ \ \ \ \ \ }%
\left( n\geq 1\right)
\end{equation}%
which makes the approximate universality of the linear trajectory slopes
explicit.

Since the AdS/CFT dictionary translates the $z$ dependence of the dilaton
into the running of the gauge coupling, the latter could be implemented into
Eq. (\ref{cnew}) for small $z\ll \Lambda _{\text{QCD}}^{-1}$ according to
the perturbative QCD $\beta $ function. This generates the leading
correction $C_{\text{pert}}(z)=-(2\ln z)^{-1}$ \cite{Kiritsis,Csaki} which{\
modifies the UV behavior of the string mode potential (\ref{vsmallz}) as $%
\mathcal{V}_{S}\left( z\right) \overset{z\rightarrow 0}{\longrightarrow }%
\frac{S^{2}-1/4}{z^{2}}+\sqrt{\frac{3}{2}}\frac{S}{z^{2}\ln (z)}+\dots $ but
vanishes in the IR. Hence asymptotic freedom and the perturbative
corrections to it can naturally coexist with confinement at large $z$.}

To summarize, we have found a new solution of the five-dimensional
Einstein-dilaton equations which provides an approximate dual background for
holographic QCD. The method used in its derivation applies to essentially
all AAdS$_{5}$ (and hence UV conformal) spacetimes with a Poincar\'{e}%
-invariant boundary. The vacuum properties of the boundary gauge theory,
including quark confinement and condensates, are dynamically encoded in this
solution without the need for additional background fields. In particular,
our background generates a confining area law for the Wilson loop (in
contrast to the soft-wall model), and we have outlined how the perturbative
running of the gauge coupling could additionally be implemented.

For metrics whose warp factors approach the power law $z^{\lambda }$ in the
infrared of the non-compact extra dimension, we found the existence
condition for the confining area law, $\lambda \geq 1$, to essentially
coincide with the condition $\lambda >1$ for an entirely discrete meson
spectrum and for the existence of a mass gap, as previously encountered in
the glueball sector. With $\lambda =2$ our background solution satisfies
this condition, and it organizes the square masses of both radial and spin
excitations into linear trajectories. It further generates the approximately
universal slope of the experimentally observed trajectories (without
introducing adjustable parameters beyond the QCD scale) and satisfactorily
reproduces the empirical spectrum of the light-flavored natural parity
mesons.

WP is grateful to C.A. Ballon Bayona, H. Boschi and G. Pimentel for fruitful
discussions. We acknowledge partial support from DAAD, CAPES, FAPESP and
CNPq.

\end{document}